\documentclass[12pt]{article}
\usepackage{amsmath,epsf,amssymb,latexsym,cite,xy}
\xyoption{matrix}\xyoption{arrow}

\newif\iffigs\figstrue

\DeclareFontFamily{U}{rsf}{}
\DeclareFontShape{U}{rsf}{m}{n}{
  <5> <6> rsfs5 <7> <8> <9> rsfs7 <10-> rsfs10}{}
\DeclareMathAlphabet\Scr{U}{rsf}{m}{n}

\def\pplogo{\vbox{\kern-\headheight\kern -29pt
\halign{##&##\hfil\cr&{%\sc
\ppnumber}\cr\rule{0pt}{2.5ex}&\ppdate\cr}
}}
\makeatletter
\def\ps@firstpage{\ps@empty \def\@oddhead{\hss\pplogo}%
  \let\@evenhead\@oddhead % in case an article starts on a left-hand page
}
\def\maketitle{\par
 \begingroup
 \def\thefootnote{\fnsymbol{footnote}}
 \def\@makefnmark{\hbox{$^{\@thefnmark}$\hss}}
 \if@twocolumn
 \twocolumn[\@maketitle]
 \else \newpage
 \global\@topnum\z@ \@maketitle \fi\thispagestyle{firstpage}\@thanks
 \endgroup
 \setcounter{footnote}{0}
 \let\maketitle\relax
 \let\@maketitle\relax
 \gdef\@thanks{}\gdef\@author{}\gdef\@title{}\let\thanks\relax}
\makeatother

\def\O{\Scr{O}}
\def\C{{\mathbb C}}
\def\P{{\mathbb P}}
\def\Q{{\mathbb Q}}
\def\R{{\mathbb R}}
\def\Z{{\mathbb Z}}

\def\Hom{\operatorname{Hom}}
\def\hom{\operatorname{hom}}

\def\Cone{\operatorname{Cone}}

\def\ch{\operatorname{\mathit{ch}}}
\def\td{\operatorname{\mathit{td}}}

\def\CY{Calabi--Yau}

\def\cI{{\Scr I}}

\def\cE{{\Scr E}}

\def\DC{\mathbf{D}}

\def\Lotimes{\mathrel{\mathop\otimes^{\mathbf{L}}}}

\begin{document}
\setcounter{page}0
\def\ppnumber{\vbox{\baselineskip14pt
\hbox{DUKE-CGTP-02-02}
\hbox{hep-th/0203111}}}
\def\ppdate{March 2002} \date{}

\title{\LARGE A Point's Point of View of Stringy Geometry\\[10mm]}
\author{
Paul S.~Aspinwall\\[2mm]
\normalsize Center for Geometry and Theoretical Physics \\
\normalsize Box 90318 \\
\normalsize Duke University \\
\normalsize Durham, NC 27708-0318
}

{\hfuzz=10cm\maketitle}

\def\Large{\large}
\def\LARGE{\large\bf}

\vskip 1cm

\begin{abstract}
The notion of a ``point'' is essential to describe the topology of
spacetime. Despite this, a point probably does not play a particularly
distinguished r\^ole in any intrinsic formulation of string theory.
We discuss one way to try to determine the notion of a point from a
worldsheet point of view. The derived category description of D-branes
is the key tool.  The case of a flop is analyzed
and $\Pi$-stability in this context is tied in to some ideas of
Bridgeland.  Monodromy associated to the flop is also computed via
$\Pi$-stability and shown to be consistent with previous conjectures.
\end{abstract}

\vfil\break

%%%%%%%%%%%%%%%%%%%%%%%%%%%%%%%%%%%%%%%%%%%%%%%%%%%%%%%%%%%%%%%%

\section{Introduction}    \label{s:intro}

One might view the central philosophy of string theory as trying to avoid
basing our interpretation of spacetime on the notion of a {\em
point}. Having said that, all our concepts of geometry do appear
ultimately to remain firmly wedded to points. It hard not to begin
any discussion of spacetime without a topological space, i.e., a
set of points.

Compactification produces an excellent laboratory for studying the
stringy geometry of space and, in particular, allows us to evade all
the thorny issues associated with time. For example, one takes a type
II string and compactifies on some \CY\ threefold $X$ to produce an
effective theory with four-dimensional spacetime. The geometry of the
\CY\ dictates the properties of the conformal field theory associated
to the non-linear sigma model of the compactification, which in turn
effects the physics in four dimensions. Thus the geometry of $X$ can
be analyzed in an inherently stringy way from the uncompactified
dimensions.  By including boundaries on the worldsheet we may analyze
D-branes (at least near the zero-coupling limit of string theory). We
can then look at ``0-branes'', i.e., D-branes which span zero
dimensions of $X$.

It is these 0-branes which form the bridge between the intrinsically
stringy notion of the worldsheet and the more classical notion of a
point. In this note we discuss a couple of properties of 0-branes to
highlight how they are key in describing the topology of the target
space.

This idea is far from new. Using 0-branes to probe the target space
has been used frequently in the past. To name just a couple of
examples, the SYZ conjecture \cite{SYZ:mir} relies on this concept, and
in \cite{DM:qiv,DGM:Dorb} it was used to analyze orbifolds.
We refer the reader to \cite{Doug:Dlect} for further discussion and references.
Reconstructing the target space using D-branes was also recently
discussed in a different context in \cite{Ber:revE}. In this note we
will discuss 0-branes in the context of the derived category point of
view \cite{Kon:mir,Doug:DC,Laz:DC,AL:DC,Dia:DC,AD:Dstab}. We do this
because we believe it is the most intrinsically ``stringy'' way of
thinking about a D-brane. The derived category point of view springs
directly from topological field theory. It is therefore superior to,
say, using an effective D-brane world-volume action which presupposes
a geometrical interpretation of the D-brane as a subspace. As the
derived category picture highlights, this latter notion is awkward in
many situations.

The basic question which emphasizes the r\^ole of 0-branes is the
following. If I gave you all the intrinsic worldsheet information
about a field theory based on a sigma model (with boundaries), could
you construct the topology of the target space? Using the ideas of
Bondal and Orlov \cite{BO:DCeq} we will discuss to what extent this is
true. Of course, the question should really be asked with ``the target
space'' replaced by ``a target space'' as it is known that frequently
two different target spaces can produce the same worldsheet
theory. The central idea is to find which D-brane-like objects can be
used as 0-branes. Unfortunately we will not find an answer to this
important question but we can give some partial criteria.

In \cite{AGM:II} it was shown that a flop between two topologically
inequivalent target spaces $X$ and $X'$ is barely noticeable from the
worldsheet point of view. As one passes through the flop, the
worldsheet theory is generically smooth even though the target space
acquires singularities. In this sense it seems much more natural to
talk about birational classes of target spaces rather than topological
classes. This is all in line with the general principle that a
conformal field theory is an algebraic construction and so algebraic
geometry should be the natural setting for stringy geometry.

In this note we will show how the behaviour of 0-branes {\em does\/}
jump discontinuously as one passes through the flop. This is based on
the ideas of Bridgeland \cite{Brig:flop} and the notion of
$\Pi$-stability \cite{DFR:stab,Douglas:D,AD:Dstab}. This issue was
also noted in \cite{Doug:DC}. In addition the flop has been analyzed from the
point of view of the D-brane world-volume in \cite{BRG:Z2}.  This
example vividly illustrates how 0-branes {\em put the topology back
into stringy geometry}.

For a particular choice of $B$ field the conformal field theory
becomes singular as one passes through the flop. It turns out that the
derived category description of the flop depends on which side of this
special point one passes. To fully understand this we will discuss
monodromy around this point. This will also produce many more
candidate 0-branes further complicating the picture.

%%%%%%%%%%%%%%%%%%%%%%%%%%%%%%%%%%%%%%%%%%%%%%%%%%%%%%%%%%%%%%%%

\section{Building a Topological Space}    \label{s:top}

The data we choose to analyze is that of the topological B-model with
boundaries. That is, we pretend that we have somehow extracted from
the worldsheet data a knowledge of all possible boundaries (consistent
with the topological field theory) and the
corresponding topological field theory of open strings stretched
between these boundaries. This data amounts to a knowledge of the
bounded derived category of coherent sheaves on $X$
\cite{Doug:DC,AL:DC}.\footnote{We are ignoring the effects of string
field theory. It is quite likely that the complete picture requires an
$A_\infty$ category as discussed in \cite{Kon:mir,Fuk:TQFT,AP:Cmir,Laz:DC}.}

Although using the derived category might at first be seen as an
unnecessary foray into obscure mathematics, it is in fact a very
natural construction from the point of view of topological field
theory. No matter how much one might try to avoid the derived category,
anyone wishing to describe the topological sector of D-branes and type
II strings would be forced to reinvent the subject or something very
similar.

The basic data we have at our disposal consists of the following:
\begin{enumerate}
\item
{\em Objects:\/} Classification of all possible topological B-type
D-branes\footnote{We refer to \cite{OOY:Dm,BDLR:Dq} for a full discussion of
what is meant exactly by ``B-type D-branes''.} associated with
worldsheet boundaries.
\item {\em Morphisms:\/} The finite-dimensional ``Hilbert Space'' of open
strings (with ghost number zero) with given D-branes at each end.
\end{enumerate}
We will use the letters $\mathsf{A},\mathsf{B},\ldots$ to refer to objects and
$\Hom(\mathsf{A},\mathsf{B})$ etc.\ to refer to the spaces of morphisms.
The joining of open strings along a common boundary gives the
multiplication law in the operator algebra for open strings. In
category language this corresponds to composition of morphisms:
\begin{equation}
\Hom(\mathsf{A},\mathsf{B})\times\Hom(\mathsf{B},\mathsf{C}) \to 
	\Hom(\mathsf{A},\mathsf{C}).
\end{equation}

As explained in \cite{Doug:DC,AL:DC}, the ghost number of the
topological field theory can be extended to the D-branes. We
may take an object $\mathsf{A}$ and produce a new object
$\mathsf{A}[n]$ by shifting the ghost numbers associated to
$\mathsf{A}$ by $n$. In terms of complexes, $\mathsf{A}[n]$ is the
complex $\mathsf{A}$ shifted, or ``translated'', left $n$ places.  The
open strings of ghost number $n$ with boundaries $\mathsf{A}$ and
$\mathsf{B}$ are then given by
$\Hom^n(\mathsf{A},\mathsf{B})=
\Hom(\mathsf{A},\mathsf{B}[n])=\Hom(\mathsf{A}[-n],\mathsf{B})$.

For the raw category, this is the only information we have. A
D-brane is just an object in the category. The only information we
have about it is the vector space of morphisms from it to any other D-brane.
How then might we decide which D-branes correspond to 0-branes?

This question was answered by Bondal and Orlov \cite{BO:DCeq} in some
cases where $X$ is not a \CY. Suppose the canonical divisor $K$ of $X$
is such that $K.C>0$ for every algebraic curve $C\subset X$, or alternatively
suppose $K.C<0$ for every algebraic curve $C\subset X$. Let us call
an $X$ satisfying either of these constraints an ``anti-\CY''.

The open strings in Witten's B-model \cite{W:CS} on which the derived
category construction is based are oriented.  In particular
$\Hom(\mathsf{A},\mathsf{B})$ need not equal
$\Hom(\mathsf{B},\mathsf{A})$. Having said that, one might try to look
for a symmetry which reverses the orientation of the strings.  This
would be given by a functor $S$ such that
\begin{equation}
\Hom(\mathsf{A},\mathsf{B}) \cong \Hom(\mathsf{B},S\mathsf{A})^*
\end{equation}
for all pairs $\mathsf{A},\mathsf{B}$. We also demand that these
isomorphisms preserve the structure of the category (we refer to
\cite{BO:DCeq} for details). Such a functor was called a ``Serre
functor'' by Bondal and Orlov and shown to be unique if it exists.

Given an algebraic variety $X$ as a target space, the
derived category $\DC(X)$ will have a Serre functor given by Serre
duality:
\begin{equation}
  S\mathsf{A} = \mathsf{A}\otimes K[d],
\end{equation}
where $d$ is the dimension of $X$.

Following Bondal and Orlov we define an object $\mathsf{P}$ in
$\DC(X)$ to be a ``BO-point'' object of codimension $s$ if there is a
Serre functor $S$ and
\begin{equation}
\begin{split}
\mathrm{i)}&\quad\,S\mathsf{P}\simeq \mathsf{P}[s]\\
\mathrm{ii)}&\quad\Hom^{<0}(\mathsf{P},\mathsf{P})=0\\
\mathrm{iii)}&\quad\Hom^0(\mathsf{P},\mathsf{P})=\C.
\end{split} \label{eq:BOpoint}
\end{equation}
The idea is that if we know $X$, then sky-scraper sheaves $\O_x$ are
indeed BO-points in $\DC(X)$. Furthermore, if $X$ is an anti-\CY\ then
all BO-point objects must be of the form $\O_x[n]$ for some $x\in X$
and $n\in\Z$. The codimension $s$ is equal to the dimension of $X$.
Thus one can identify the points of $X$ purely from their categorical
description. This identification will fail for the case of \CY's but
for the meantime let us assume that we have determined these ``point
objects'' for $X$.

Not only can we determine the set of points on $X$ but we may also
determine $X$ as a topological space. To do this it is necessary to describe
the set of open sets on $X$. The first step in this process is to
identify elements of $\DC(X)$ which correspond to (a complex with a
single nonzero term given by) a line bundle. If we have already
identified the points as above then this is easy. An object
$\mathsf{L}$ is a ``line-bundle object'' if there is an $s\in\Z$
(which may depend on $\mathsf{P}$) such that
\begin{equation}
\begin{split}
\mathrm{i)}&\quad\Hom^s(\mathsf{L},\mathsf{P})=\C\\
\mathrm{ii)}&\quad\Hom^t(\mathsf{L},\mathsf{P})=0\quad\hbox{for $t\neq s$,}
\end{split} \label{eq:LBobj}
\end{equation}
for all point objects $\mathsf{P}$.
Again Bondal and Orlov showed that if $X$ exists then such line bundle
objects correspond exactly to line bundles possibly translated in the
complex by an arbitrary integer.

It is useful to fix the arbitrary translation symmetry in
$\DC(X)$. Choose a point object $\mathsf{P}_0$. Next choose any line
bundle object $\mathsf{L}_0$ which satisfies (\ref{eq:LBobj}) for
$s=0$ with respect to $\mathsf{P}_0$. A point object $\mathsf{P}$ will
then be called an ``absolute'' point object if
$\Hom^0(\mathsf{L}_0,\mathsf{P})=\C$. This effectively restricts
attention to point objects corresponding to sky-scraper sheaves at a
fixed position in the complex. The set of absolute point objects is
now in one-to-one correspondence with points on $X$.  One may similarly
define ``absolute'' line bundle objects as line bundle objects
$\mathsf{L}$ satisfying $\Hom^0(\mathsf{L},\mathsf{P}_0)=\C$.

Consider now a morphism $\alpha\in\Hom(\mathsf{L}_1,\mathsf{L}_2)$ for
an arbitrary pair $\mathsf{L}_1,\mathsf{L}_2$ of absolute line bundle
objects. This induces a map
\begin{equation}
  \alpha^*_{\mathsf{P}}:\Hom(\mathsf{L}_2,\mathsf{P})\to
	\Hom(\mathsf{L}_1,\mathsf{P})
\end{equation}
for any object $\mathsf{P}$ in the obvious way. Let $U_\alpha$ be the
set of absolute point objects in $\DC(X)$ such that
$\alpha^*_{\mathsf{P}}\neq0$.  The geometrical meaning of $U_\alpha$
is as follows. $\alpha$ is a global map from $\mathsf{L}_1$ to
$\mathsf{L}_2$ and is thus equivalent to a section of
$\mathsf{L}_1^\vee\otimes \mathsf{L}_2$. A divisor $D_\alpha$ can be
associated to the zeroes of this section in the usual way. If $\alpha$
is chosen suitably, this divisor corresponds to a codimension one
subspace of $X$. $U_\alpha$ is then the open set in $X$ given be the
complement of $D_\alpha$. These open sets $U_\alpha$ can be used to generate
a topology (namely the Zariski topology) for $X$.

%%%%%%%%%%%%%%%%%%%%%%%%%%%%%%%%%%%%%%%%%%%%%%%%%%%%%%%%%%%%%%%%%

\section{Hunting the 0-Brane}  \label{s:D0}

If we could do the same when $X$ is a \CY\ then we would be done.  The
Serre functor in this case is given by $S\mathsf{A}=\mathsf{A}[d]$,
where $d$ is the dimension of $X$. This may be seen directly at the
worldsheet level. One can change the orientation of the open strings
by reversing the sign of the ghost number. Then use a spectral flow
argument along the lines of \cite{LVW:} to shift the ghost numbers by
$d$. This gives the desired Serre functor as claimed.

The problem for the \CY\ case is that condition i) in
(\ref{eq:BOpoint}) is trivial thus weakening the criteria. There are
many objects in $\DC(X)$ which satisfy the BO-point condition without
being sky-scraper sheaves. We therefore need a more stringent
condition for being called a ``point object''. If $x$ and $y$ are
points in $X$ then one may show that\footnote{For this and many
subsequent calculations the following spectral sequence is useful. Let
$i:Z\to X$ be a smooth embedding with $N$ the normal bundle. Then if
$\Scr{A},\Scr{B}$ are sheaves on $Z$
\begin{equation}
  E_2^{p,q}=\Hom_{\O_Z}^p(\Scr{A},\Scr{B}\otimes\Lambda^qN)\Rightarrow
	\Hom_{\DC(X)}^{p+q}(i_*\Scr{A},i_*\Scr{B}).  \label{eq:Bss}
\end{equation}}
\begin{equation}
\begin{split}
  \dim\Hom^q(\O_x,\O_x) &= \binom dq\\
  \Hom^q(\O_x,\O_y) &= 0 \quad\hbox{for $x\neq y$.}
\end{split}  \label{eq:point}
\end{equation}
This motivates the following.  An object $\mathsf{A}$ in $\DC(X)$
satisfying $\dim\Hom^q(\mathsf{A},\mathsf{A})=\binom dq$ will be
called a ``pre-point object''.  Two non-isomorphic pre-point objects
$\mathsf{A}$ and $\mathsf{B}$ will be called ``mutually consistent''
if they satisfy $\dim\Hom^q(\mathsf{A},\mathsf{B})=0$ for any $q$.

Unfortunately there are many more pre-point objects in $\DC(X)$ than
there are points. It is not hard to see in any example that many pairs of
pre-point objects will not be consistent in the above sense. Let us
consider how we can further restrict the conditions on a object being
declared a point. 

As an example consider the case where $X$ is a quintic hypersurface in
$\P^4$. Monodromy around loops in the moduli space of complexified
K\"ahler forms is believed to induce autoequivalences of $\DC(X)$ in
accordance with Kontsevich's homological mirror symmetry
conjecture. See \cite{Kon:mir,Horj:DX,me:navi}, for example, for an
account of this. In particular monodromy around the ``conifold'' point
in the case of the quintic is obtained from an autoequivalence given
by a particular Fourier--Mukai transform
\cite{Kont:mon,Horj:DX,ST:braid}.
Under such a transform, the sky-scraper sheaf of a point $\O_x$ is
transformed into $\cI_x[1]$ --- the ideal sheaf of functions vanishing
at the same point which is then shifted one place left in the complex.

It is therefore impossible to declare that $\O_x$ has any more right
to be associated to a point than $\cI_x[1]$ when only looking at
intrinsic categorical data. Lack of mutual consistency can be seen
from the fact that $\Hom^1(\cI_x[1],\O_y)=\C$ for any $y\neq x$.

One possibility which should immediately spring to mind as regards
fixing this problem is that of ``D-brane charge''. We may restrict our
collection of candidate points by insisting that they all have the
same D-Brane charge.

For the derived category of coherent sheaves, an
object 
\begin{equation}
  \ldots\to\cE^0\to\cE^1\to\cE^2\to\ldots,
\end{equation}
will have a D-brane charge given by the Chern
character\footnote{D-brane charge is measured by K-theory. By
considering the Chern character we are only looking at the free
part. The torsion part is accessible from the derived category and
might be of interest in some cases.} 
\begin{equation}
  \ch(\cE^\bullet) = \sum_n(-1)^n\ch(\cE^n).
\end{equation}
The intrinsic categorical information knows about the Chern character
because of the Hirze\-bruch-Riemann-Roch theorem:
\begin{equation}
  \sum_n(-1)^n\dim\Hom(\mathsf{A},\mathsf{B}[n])=
     \int_X\ch(\mathsf{A})\ch(\mathsf{B})^\vee\td(T_X).
\end{equation}
This gives a skew-symmetric inner product on the derived category
which is believed to be mirror to the intersection form of 3-cycles
\cite{HM:alg2,BDLR:Dq,HIV:D-mir}. As such it should be non-degenerate. Also,
assuming the Hodge conjecture to be true, the Chern characters of all
objects in $\DC(X)$ span $H^{\textrm{even}}(X,\Q)$. We may therefore
claim the following: If $\mathsf{A}$ and $\mathsf{A}'$ are two objects
in $\DC(X)$ and
\begin{equation}
  \sum_n(-1)^n\dim\Hom(\mathsf{A},\mathsf{B}[n]) = 
	\sum_n(-1)^n\dim\Hom(\mathsf{A}',\mathsf{B}[n]),
\end{equation}
for all objects $\mathsf{B}$ in $\DC(X)$, then $\mathsf{A}$ and
$\mathsf{A}'$ have the same Chern character and therefore D-brane
charge. Thus we know how to analyze D-brane charge from purely
categorical data.

The approach one might try to take could then be as follows. Choose a
pre-point object in $\DC(X)$. Now catalogue all other pre-point
objects in $\DC(X)$ having the same D-brane charge. In simple cases it
is perhaps conceivable  that one might obtain a {\em mutually
consistent\/} set of pre-point objects. The pre-point objects would
hopefully therefore be fully accounted for by all the points on $X$
and their translations in $\DC(X)$ by an even number. Unfortunately
this procedure is doomed to failure in some (if not all) cases as we
now demonstrate.

%%%%%%%%%%%%%%%%%%%%%%%%%%%%%%%%%%%%%%%%%%%%%%%%%%%%%%%%%%%%%%

\section{The Flop}   \label{s:flop}

Suppose $X$ contains a rational curve $C$ which may be contracted down
to a point by varying the K\"ahler form. We may then ``proceed
through'' this wall of the K\"ahler cone to produce a flop of $X$ into
another \CY\ space $X'$ with a new corresponding rational curve
$C'$. In general $X'$ can be expected to be topologically distinct
from $X$.

The topological $B$-model is insensitive to any change in the K\"ahler
form and so $\DC(X)$ must remain invariant under such a process. 
Indeed, it was shown explicitly in \cite{BO:flop,Brig:flop} that
$\DC(X)$ is equivalent to $\DC(X')$. This
presents a challenge to our program. If we are to construct $X$ from
$\DC(X)$ we should equally be allowed to construct $X'$.

\iffigs
\begin{figure}
  \centerline{\epsfxsize=8cm\epsfbox{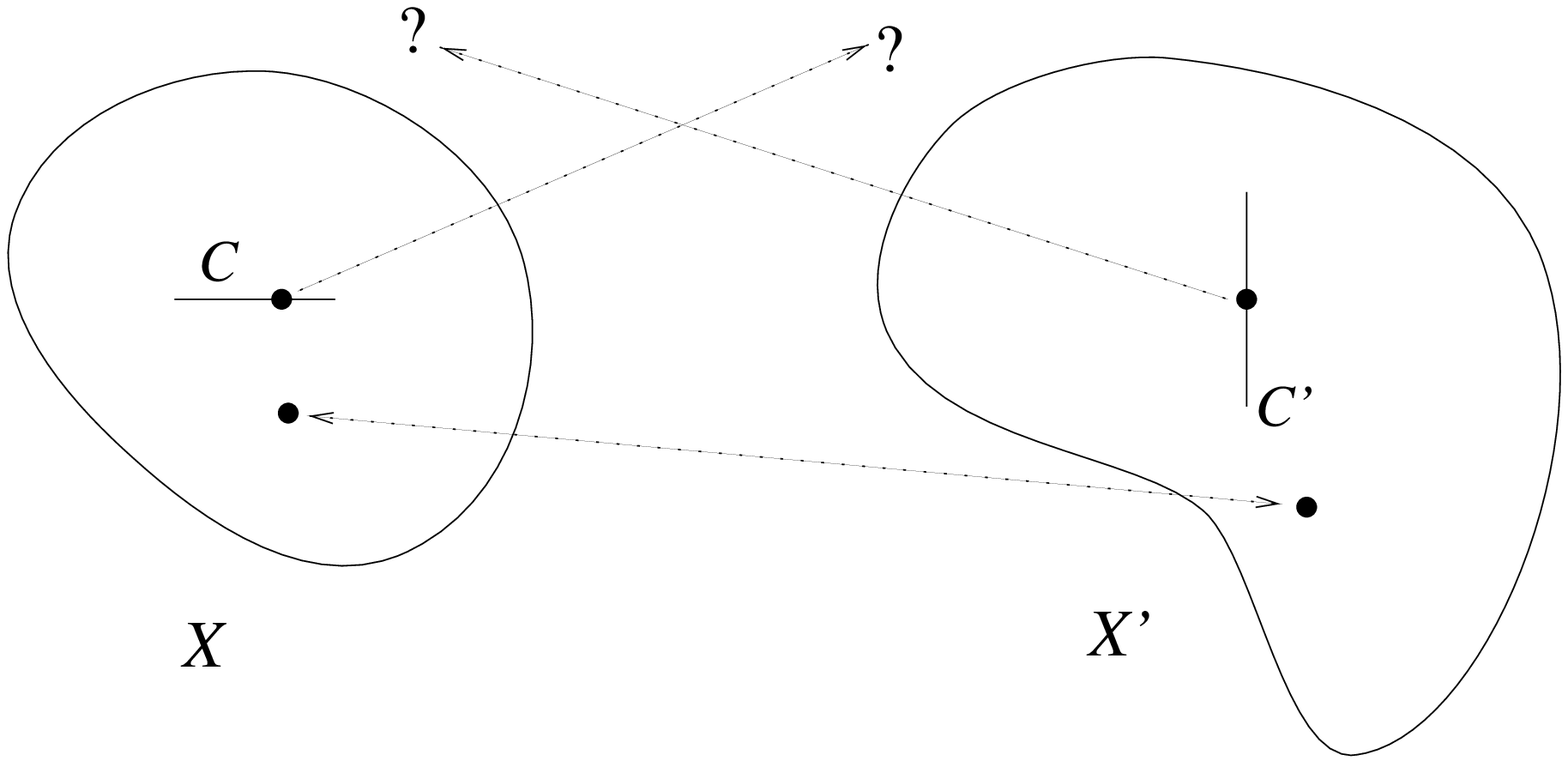}}
  \caption{The correspondence between points in $X$ and $X'$.}
  \label{f:cor}
\end{figure}
\fi

Bridgeland \cite{Brig:flop} gave an explicit form to the equivalence
between $\DC(X)$ and $\DC(X')$. Away from the curves $C$
and $C'$, the spaces $X$ and $X'$ are identical. That is, we have a
one-to-one correspondence between points in these two spaces as shown
schematically in figure~\ref{f:cor}. Naturally this correspondence
breaks down for points within $C$ and $C'$. A sky-scraper sheaf on
a point in $C$ corresponds to an object in $\DC(X')$ which is not
simply a sheaf but rather a longer complex.\footnote{In fact it's a
``perverse sheaf'' \cite{Brig:flop}.} Similarly a point $x'\in C'$
corresponds to a specific non-sheaf object $\mathsf{Z}_{x'}$ in $\DC(X)$.

This object $\mathsf{Z}_{x'}$ is clearly a pre-point object and has the
same D-brane charge as a point object in $X$. It does not however
correspond to a point in $X$. We therefore need more information to
rule it out as a point in $X$. Since the topological field theory
cannot distinguish between $X$ and $X'$, we are forced to go beyond the
topological field theory, and thus the intrinsic information content
of the derived category.

We want to use the $\Pi$-stability of \cite{DFR:stab} as a criterion
for ruling out $\mathsf{Z}_{x'}$. In the region of K\"ahler form moduli space
corresponding roughly to the K\"ahler cone of $X$ we will show that
$\mathsf{Z}_{x'}$ is unstable. This was also alluded to in \cite{Doug:DC}.

The data required for $\Pi$-stability is a ``grading'' $\varphi\in\R$
associated to a subset of the objects in $\DC(X)$. This subset of
objects is the $\Pi$-stable set of objects. The grading is determined
mod 2 as the phase of the central charge of the corresponding D-brane
and, as such, depends continuously on the complexified K\"ahler form
$B+iJ$.

Note that the central charge of the D-brane is data visible in the
non-compact 4-dimensional $N=2$ supersymmetric field theory and thus
fits into our game rules. It is data we are allowed to use to
construct $X$ without knowing anything about $X$ beforehand.

The exact rules for $\Pi$-stability are rather convoluted as explained
in \cite{AD:Dstab}. As $B+iJ$ is varied, objects will enter and leave
the set of stable objects according to their values of $\varphi$. The
statement of $\Pi$-stability is rather simple if we focus attention on
a basic set of objects $\{\mathsf{A},\mathsf{B},\mathsf{C}\}$ where we
know $\mathsf{C}$ can potentially decay to $\bar{\mathsf{A}}$ (i.e., the
anti-brane of $\mathsf{A}$ which is actually the same thing as
$\mathsf{A}[1]$ as argued in \cite{Doug:DC}) and $\mathsf{B}$. Assume
$\bar{\mathsf{A}}$ and $\mathsf{B}$ are stable and we wish to determine
if $\mathsf{C}$ is stable relative to decay into $\bar{\mathsf{A}}$ and
$\mathsf{B}$.

Such a triple of objects forms a ``distinguished triangle'' in
$\DC(X)$ and is written:
\begin{equation}
\xymatrix{
&\mathsf{C}\ar[dl]|{[1]}&\\
\mathsf{A}\ar[rr]&&\mathsf{B}.\ar[ul]
} \label{eq:triABC}
\end{equation}
The ``[1]'' denotes that the morphism is really from $\mathsf{C}$ to
$\mathsf{A}[1]$. $\mathsf{C}$ is then stable with respect to
$\mathsf{A}$ and $\mathsf{B}$ if and only if
$\varphi(\mathsf{B})-\varphi(\mathsf{A})<1$.

It is important to note that the ``[1]'' in (\ref{eq:triABC}) can be
shuffled around to any edge. For example, (\ref{eq:triABC}) can be
rewritten
\begin{equation}
\xymatrix{
&\mathsf{C}\ar[dl]&\\
\mathsf{A}[1]\ar[rr]|{[1]}&&\mathsf{B}.\ar[ul]
} 
\end{equation}
This symmetry between the edges means that if the difference in
$\varphi$'s exceeds one on any of the edges of the triangle, the
object in the opposite vertex will decay. This symmetry is one of the important
properties of the derived category which cannot be reproduced in any
``abelian'' category such as the category of vector bundles, and shows
why categories such as the latter cannot fully model D-brane decay
\cite{AD:Dstab}.

In order to analyze the flop we use the following D-branes as building
blocks:
\begin{enumerate}
\item The skyscraper sheaf $\O_x$ for $x\in C$.
\item The skyscraper sheaf $\O_y$ for $y\not\in C$.
\item The structure sheaf $\O_C$ of the flopping curve, i.e., the
2-brane wrapped around $C$.
\end{enumerate}
The periods of these D-branes are very easy to compute. We can use
mirror symmetry to compute periods exactly (with respect to
$\alpha'$-corrections) as in \cite{DG:fracM,AD:Dstab}. We may localize
the picture by assuming that all curves in $X$ are very large except
for those homologous to $C$. The required Picard--Fuch's differential
equation was then determined in section 5.4 of \cite{AGM:sd}.

\iffigs
\begin{figure}
  \centerline{\epsfxsize=5cm\epsfbox{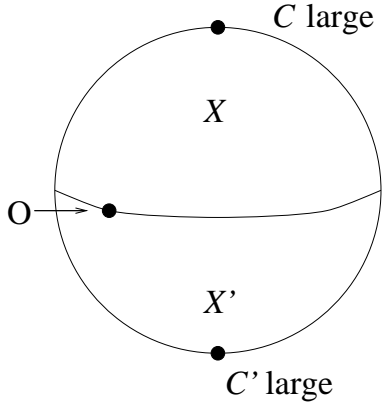}}
  \caption{The moduli space of $B+iJ$ for the flop.}
  \label{f:BJs}
\end{figure}
\fi

The part of the moduli space of $B+iJ$ which concerns us can be
viewed as a 2-sphere as shown in figure \ref{f:BJs}. The large radius
limit of $C\subset X$ is put at one pole and the large radius limit of
$C'\subset X'$ is put at the other pole. On the equator, which
separates the $X$ ``phase'' from the $X'$ ``phase'' we have a point
where the worldsheet theory becomes singular. This latter point will be
viewed as the origin in our final coordinate system, so we label it
$O$.

There is a natural ``algebraic'' coordinate $z$ on this sphere with
$z=0$ in the large $C$ limit, $z=\infty$ in the large $C'$ limit and
$z=1$ at $O$. In terms of this coordinate $z$ we have a Picard--Fuchs
equation
\begin{equation}
  \left(z\frac{d}{dz}\right)^2\Phi - z\left(z\frac{d}{dz}\right)^2\Phi=0.
	\label{eq:PFflop}
\end{equation}
This has a general solution
\begin{equation}
  \Phi=C_1+C_2\log(z).
\end{equation}
The rules of the ``mirror map'' \cite{AGM:mdmm} (see \cite{CK:mbook}
for more details) then tell us
\begin{equation}
  t = \int_C B+iJ = \frac1{2\pi i}\frac{\Phi_1}{\Phi_0},
\end{equation}
where $\Phi_0$ is the period which is regular and equal to 1 at the
large $C$ limit (i.e., $z\to0$ limit) and $\Phi_1$ is the period whose
limiting behaviour is $\log(z)$ for $z\to0$. Clearly from
(\ref{eq:PFflop}) we have
\begin{equation}
  t = \frac1{2\pi i}\log(z).
\end{equation}
Thus $t=0$ at $O$ as promised.

In order to compute central charges we use
\cite{MM:K,FW:D}\footnote{Note that we are using the sign convention
for $B$ from section 3 rather than section 2 of \cite{AD:Dstab}!}
\begin{equation}
Z(\mathsf{A}) = \int_X
        e^{B+iJ}\ch(\mathsf{A})\sqrt{\td(T_X)}+\hbox{quantum
        corrections} \label{eq:LRZ}
\end{equation}
where the quantum corrections are determined from the fact that
$Z(\mathsf{A})$ is a period. This immediately yields:
\begin{equation}
\begin{split}
Z(\O_x)=Z(\O_y)&=1\\
Z(\O_C)&=t,
\end{split}
\end{equation}
i.e., there are no quantum corrections!
This makes the flop particularly easy to analyze. In most other examples
the complexity of the Picard--Fuchs system makes the formula for
central charged much less amenable as seen in \cite{AD:Dstab}.

\iffigs
\begin{figure}
  \centerline{\epsfxsize=6cm\epsfbox{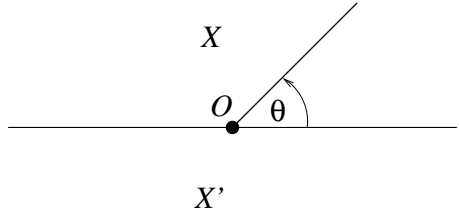}}
  \caption{The complex $t$-plane.}
  \label{f:theta}
\end{figure}
\fi

We may capture all the relevant properties of D-brane behaviour near
the flop by assuming $t=\epsilon e^{i\theta}$ where
$\epsilon\ll1$. We thus parametrize the moduli space as in
figure~\ref{f:theta}. 

The grades of our D-branes can be determined from $\varphi=-\frac1\pi
\arg(Z)$ where at large radius limit (i.e., $\theta=\frac\pi2$) we
insist that $-2<\varphi\leq0$ 
\cite{AD:Dstab}. Thus
\begin{equation}
\begin{split}
\varphi(\O_x)=\varphi(\O_y)&=0\\
\varphi(\O_C)&=-\frac\theta\pi.
\end{split}
\end{equation}

We may now compute D-brane stability. The first distinguished triangle
of interest is
\begin{equation}
\xymatrix{
&\mathsf{A}\ar[dl]_(0.3){1-\frac\theta\pi}|{[1]}&\\
\O_C\ar[rr]_{\frac\theta\pi}&&\O_x\ar[ul]_0
} \label{eq:triA}
\end{equation}
Here we label each edge of the triangle with the difference in
grading between the vertices. It is whether this label is greater than
or less than one which determines the stability of the opposite
vertex. In derived category language $\mathsf{A}$ is the ``Cone'' of the natural
map $\O_C\to\O_x$. The short exact sequence
\begin{equation}
0\to\O_C(-1)\to\O_C\to\O_x\to0 \label{eq:ses1}
\end{equation}
tells us that $\mathsf{A}$ is given by $\O_C(-1)[1]$. The triangle
(\ref{eq:triA}) dictates that $\O_x$ will decay into $\mathsf{A}$ and
$\O_C$ if $\theta<0$. That is the 0-brane $\O_x$ becomes unstable as
we pass from the $X$ phase into the $X'$ phase through
$\theta=0$. {\em This is exactly what we wanted!} If we impose the
stability constraint on our 0-branes then we lose the ``bogus'' ones
living on a curve flopped out of existence.

Note that there is {\em no\/} nontrivial homomorphism $\O_C\to\O_y$ with
$y\not\in C$. This means that there is no distinguished triangle of
the form (\ref{eq:triA}) with $x$ replaced by $y$. This prevents the
0-branes $\O_y$, away from the flop, from decaying --- again in line with
expectation.

We would now like to find the new 0-branes which jump into existence
as we pass into the $X'$ phase. It would be nice to have a systematic
way of building triangles relevant to our discussion. Clearly
homomorphisms from $\O_C$ to something else will be of interest since
the relative grading is likely to change as we encircle $O$. We might
therefore consider a map $\O_C\to\mathsf{A}$ next. Using the spectral
sequence (\ref{eq:Bss}) we can show that $\Hom(\O_C,\mathsf{A})=0$ so
we get nothing new here. As a next attempt one can show that
$\Hom(\O_C[-1],\mathsf{A})=\C^2$. Let us see what happens
when we consider the cone $\mathsf{B}$ of a map
$\O_C[-1]\to\mathsf{A}$. This generates the triangle
\begin{equation}
\xymatrix{
&\mathsf{B}\ar[dl]_(0.4){-\frac\theta\pi}|{[1]}&\\
\O_C[-1]\ar[rr]_{1+\frac\theta\pi}&&\mathsf{A}\ar[ul]_0
} \label{eq:triB}
\end{equation}
Thus $\mathsf{B}$ jumps into existence as $\theta$ becomes negative,
i.e., we pass into the $X'$ phase. It is not hard to show that $B$ is
a pre-point object in the sense of (\ref{eq:point}) and has the same
D-brane charge as a point. Since
$\Hom(\O_C[-1],\mathsf{A})=\C^2$ we had a choice of which map to use
to give $\mathsf{B}$. This essentially gives us a whole $\P^1$'s-worth
of objects $\mathsf{B}$.  Indeed $\mathsf{B}$ is exactly the object in
$\DC(X)$ studied by Bridgeland in \cite{Brig:flop} where he showed
that the $\P^1$ parametrized by the $\mathsf{B}$ objects fits nicely
into $X-C$ to produce precisely $X'$. That is, he showed rigorously
that the objects $\mathsf{B}$ really are the point objects for points
on $C'\subset X'$, and so the objects $\mathsf{B}$ are the $\mathsf{Z}_{x'}$
objects discussed earlier in this section.

It is very gratifying to note that the objects selected by Bridgeland
for this purpose are exactly the ones selected by $\Pi$-stability for
this same r\^ole.

Let us now continue this process. Since $\Hom(\O_C[-1],\mathsf{B})=\C$, we may
construct another distinguished triangle:
\begin{equation}
\xymatrix{
&\mathsf{C}\ar[dl]_(0.4){-\frac\theta\pi}|{[1]}&\\
\O_C[-1]\ar[rr]_{1+\frac\theta\pi}&&\mathsf{B},\ar[ul]_0
}
\end{equation}
which shows that another D-brane, $\mathsf{C}$, comes into existence along with
$B$ as we pass into the $X'$ phase. Lastly $\Hom(\O_C[-1],\mathsf{C})=0$ but
$\Hom(\O_C[-2],\mathsf{C})=\C^2$ so the triangle
\begin{equation}
\xymatrix{
&\mathsf{D}\ar[dl]_(0.4){-1-\frac\theta\pi}|{[1]}&\\
\O_C[-2]\ar[rr]_{2+\frac\theta\pi}&&\mathsf{C}\ar[ul]_0
} 
\label{eq:triD}
\end{equation}
brings a new object $\mathsf{D}$ into existence as $\theta$ falls
below $-\pi$, i.e., after we circle $O$ clockwise in
figure~\ref{f:theta} and come back into the $X$ phase. 
Note that since $\Hom(\O_C[-2],\mathsf{C})=\C^2$, we actually have a
$\P^1$'s-worth of objects $\mathsf{D}$. We claim that the $\mathsf{D}$
objects are the new objects representing points on $C$ after monodromy
around $O$. This is exactly the mechanism for monodromy presented in
the context of the quintic in \cite{AD:Dstab} except that here we have
had to pass through a number of triangles to perform the
monodromy. Note also that even though $\mathsf{D}$ and $\O_x$ are
quite different objects in $\DC(X)$, they have the same D-brane
charge. Thus this monodromy cannot be seen purely at the level of
cohomology classes.

Note that this monodromy is consistent with the Fourier--Mukai
transform predicted by Horja
\cite{Horj:DX,Horj:EZ} and Seidel and Thomas \cite{ST:braid}. That is,
in the language of \cite{ST:braid}:
\begin{equation}
  \mathsf{D} = \Cone(\hom(\O_C,\O_x)\Lotimes\O_C\to\O_x).
\end{equation}

In order to further clarify this monodromy statement, let us consider
what happens if we try to perform the flop by moving {\em
counter-clockwise\/} around $O$ to leave the $X$ phase and pass into
the $X'$ phase. Now we have the following relevant distinguished
triangle
\begin{equation}
\xymatrix{
&\O_x\ar[dl]_(0.4){1-\frac\theta\pi}|{[1]}&\\
\O_C\ar[rr]_{\frac\theta\pi}&&\O_C(1),\ar[ul]_0
} 
\label{eq:tricc}
\end{equation}
which shows how $\O_x$ becomes unstable as $\theta>\pi$.
We claim then the triangle (\ref{eq:triD}) is exactly the triangle
(\ref{eq:tricc}) under clockwise monodromy around $O$, i.e., $\O_C$
becomes $\O_C[-2]$ and $\O_C(1)$ becomes $\mathsf{C}$. This is again
consistent with the Fourier--Mukai transform above.

%%%%%%%%%%%%%%%%%%%%%%%%%%%%%%%%%%%%%%%%%%%%%%%%%%%%%%%%%%%%%%%%%%%

\section{Discussion}  \label{s:conc}

The derived category picture emphasizes the issues involved in
declaring a given D-brane to be a point, i.e., a 0-brane. The
occurrence of flops and monodromy show that many objects in the derived
category have a right to be called a point, but only in a particular
region of K\"ahler moduli space. 

In any sufficiently complex example it is clear that even if one
restricts to a specific D-brane charge, there are an infinite number
of candidate point objects for each point in $X$. Many of these are
related by monodromy transforms like the one we saw above for a
flop. One can also generate similar monodromy transforms for points on
the exceptional divisor of an orbifold.
Insisting that a point object be {\em stable\/} removes all the
monodromy images of each point.

It is overly-optimistic to assume that we have given enough conditions
to cut down the number of pre-point objects to be left with the
consistent set of true point objects. Understanding this further is
key to knowing how to extract target space topology from worldsheet
information.

%%%%%%%%%%%%%%%%%%%%%%%%%%%%%%%%%%%%%%%%%%%%%%%%%%%%%%%%%%%%%%%%%%%

\section*{Acknowledgments}

It is a pleasure to thank T.~Bridgeland, S.~Katz, D.~Morrison and R.~Plesser
for useful conversations.  The author is supported in part by NSF grant
DMS-0074072 and by a research fellowship from the Alfred P.~Sloan
Foundation.

%\bibliographystyle{my-phys}
%\bibliography{string}

\begin{thebibliography}{10}

\bibitem{SYZ:mir}
A.~Strominger, S.-T. Yau, and E.~Zaslow,
\newblock {\em Mirror Symmetry is T-Duality},
\newblock Nucl. Phys. {\bf B479} (1996) 243--259, hep-th/9606040.

\bibitem{DM:qiv}
M.~R. Douglas and G.~Moore,
\newblock {\em D-branes, Quivers, and ALE Instantons},
\newblock hep-th/9603167.

\bibitem{DGM:Dorb}
M.~R. Douglas, B.~R. Greene, and D.~R. Morrison,
\newblock {\em Orbifold Resolution by D-Branes},
\newblock Nucl. Phys. {\bf B506} (1997) 84--106, hep-th/9704151.

\bibitem{Doug:Dlect}
M.~R. Douglas,
\newblock {\em Two Lectures on D-Geometry and Noncommutative Geometry},
\newblock in M.~Duff et~al., editors, ``Nonperturbative Aspects of Strings,
  Branes and Supersymmetry'', pages 131--156, World Scientific, 1999,
\newblock hep-th/9901146.

\bibitem{Ber:revE}
D.~Berenstein,
\newblock {\em Reverse Geometric Engineering of Singularities},
\newblock hep-th/0201093.

\bibitem{Kon:mir}
M.~Kontsevich,
\newblock {\em Homological Algebra of Mirror Symmetry},
\newblock in ``Proceedings of the International Congress of Mathematicians'',
  pages 120--139, Birkh{\"a}user, 1995,
\newblock alg-geom/9411018.

\bibitem{Doug:DC}
M.~R. Douglas,
\newblock {\em D-Branes, Categories and $N$=1 Supersymmetry},
\newblock J. Math. Phys. {\bf 42} (2001) 2818--2843, hep-th/0011017.

\bibitem{Laz:DC}
C.~I. Lazaroiu,
\newblock {\em Unitarity, D-Brane Dynamics and D-brane Categories},
\newblock JHEP {\bf 12} (2001) 031, hep-th/0102183.

\bibitem{AL:DC}
P.~S. Aspinwall and A.~E. Lawrence,
\newblock {\em Derived Categories and Zero-Brane Stability},
\newblock JHEP {\bf 08} (2001) 004, hep-th/0104147.

\bibitem{Dia:DC}
D.-E. Diaconescu,
\newblock {\em Enhanced D-brane Categories from String Field Theory},
\newblock JHEP {\bf 06} (2001) 016, hep-th/0104200.

\bibitem{AD:Dstab}
P.~S. Aspinwall and M.~R. Douglas,
\newblock {\em D-Brane Stability and Monodromy},
\newblock hep-th/0110071.

\bibitem{BO:DCeq}
A.~Bondal and D.~Orlov,
\newblock {\em Reconstruction of a Variety from the Derived Category and Groups
  of Autoequivalences},
\newblock alg-geom/9712029.

\bibitem{AGM:II}
P.~S. Aspinwall, B.~R. Greene, and D.~R. Morrison,
\newblock {\em \CY\ Moduli Space, Mirror Manifolds and Spacetime Topology
  Change in String Theory},
\newblock Nucl. Phys. {\bf B416} (1994) 414--480.

\bibitem{Brig:flop}
T.~Bridgeland,
\newblock {\em Flops and Derived Categories},
\newblock math.AG/0009053.

\bibitem{DFR:stab}
M.~R. Douglas, B.~Fiol, and C.~R{\"o}melsberger,
\newblock {\em Stability and BPS Branes},
\newblock hep-th/0002037.

\bibitem{Douglas:D}
M.~R. Douglas,
\newblock {\em D-Branes on Calabi-Yau Manifolds},
\newblock math.AG/0009209.

\bibitem{BRG:Z2}
B.~R. Greene,
\newblock {\em D-brane Topology Changing Transitions},
\newblock Nucl. Phys. {\bf B525} (1998) 284--296, hep-th/9711124.

\bibitem{Fuk:TQFT}
K.~Fukaya and P.~Seidel,
\newblock {\em Floer Homology, $A_\infty$-Categories and Topological Field
  Theory},
\newblock Lecture Notes in Pure and Appl. Math. {\bf 184} (1997) 9--32.

\bibitem{AP:Cmir}
A.~Polishchuk and E.~Zaslow,
\newblock {\em Categorical Mirror Symmetry: The Elliptic Curve},
\newblock Adv. Theor. Math. Phys. {\bf 2} (1998) 443--470, math.AG/9801119.

\bibitem{OOY:Dm}
H.~Ooguri, Y.~Oz, and Z.~Yin,
\newblock {\em D-branes on Calabi-Yau Spaces and their Mirrors},
\newblock Nucl. Phys. {\bf B477} (1996) 407--430, hep-th/9606112.

\bibitem{BDLR:Dq}
I.~Brunner, M.~R. Douglas, A.~Lawrence, and C.~R{\"o}melsberger,
\newblock {\em D-branes on the Quintic},
\newblock JHEP {\bf 08} (2000) 015, hep-th/9906200.

\bibitem{W:CS}
E.~Witten,
\newblock {\em Chern-Simons Gauge Theory as a String Theory},
\newblock in H.~Hofer et~al., editors, ``The Floer Memorial Volume'', pages
  637--678, Birkh{\"a}user, 1995,
\newblock hep-th/9207094.

\bibitem{LVW:}
W.~Lerche, C.~Vafa, and N.~P. Warner,
\newblock {\em Chiral Rings in $N=2$ Superconformal Theories},
\newblock Nucl. Phys. {\bf B324} (1989) 427--474.

\bibitem{Horj:DX}
R.~P. Horja,
\newblock {\em Hypergeometric Functions and Mirror Symmetry in Toric
  Varieties},
\newblock math.AG/9912109.

\bibitem{me:navi}
P.~S. Aspinwall,
\newblock {\em Some Navigation Rules for D-brane Monodromy},
\newblock J. Math. Phys. {\bf 42} (2001) 5534--5552, hep-th/0102198.

\bibitem{Kont:mon}
M.~Kontsevich, 1996,
\newblock Rutgers Lecture, unpublished.

\bibitem{ST:braid}
P.~Seidel and R.~P. Thomas,
\newblock {\em Braid Groups Actions on Derived Categories of Coherent Sheaves},
\newblock Duke Math. J. {\bf 108} (2001) 37--108, math.AG/0001043.

\bibitem{HM:alg2}
J.~A. Harvey and G.~Moore,
\newblock {\em On the algebras of BPS states},
\newblock Commun. Math. Phys. {\bf 197} (1998) 489--519, hep-th/9609017.

\bibitem{HIV:D-mir}
K.~Hori, A.~Iqbal, and C.~Vafa,
\newblock {\em D-branes and Mirror Symmetry},
\newblock hep-th/0005247.

\bibitem{BO:flop}
A.~Bondal and D.~Orlov,
\newblock {\em Semiorthogonal Decomposition for Algebraic Varieties},
\newblock alg-geom/9506012.

\bibitem{DG:fracM}
D.-E. Diaconescu and J.~Gomis,
\newblock {\em Fractional Branes and Boundary States in Orbifold Theories},
\newblock JHEP {\bf 10} (2000) 001, hep-th/9906242.

\bibitem{AGM:sd}
P.~S. Aspinwall, B.~R. Greene, and D.~R. Morrison,
\newblock {\em Measuring Small Distances in $N=2$ Sigma Models},
\newblock Nucl. Phys. {\bf B420} (1994) 184--242, hep-th/9311042.

\bibitem{AGM:mdmm}
P.~S. Aspinwall, B.~R. Greene, and D.~R. Morrison,
\newblock {\em The Monomial-Divisor Mirror Map},
\newblock Internat. Math. Res. Notices {\bf 1993} 319--338, alg-geom/9309007.

\bibitem{CK:mbook}
D.~A. Cox and S.~Katz,
\newblock {\em Mirror Symmetry and Algebraic Geometry}, Mathematical Surveys
  and Monographs~{\bf 68},
\newblock AMS, 1999.

\bibitem{MM:K}
R.~Minasian and G.~Moore,
\newblock {\em K-Theory and Ramond-Ramond Charge},
\newblock J. High Energy Phys. {\bf 11} (1997) 002, hep-th/9710230.

\bibitem{FW:D}
D.~S. Freed and E.~Witten,
\newblock {\em Anomalies in String Theory with D-branes},
\newblock hep-th/9907189.

\bibitem{Horj:EZ}
P.~Horja,
\newblock {\em Derived Category Automorphisms from Mirror Symmetry},
\newblock math.AG/0103231.

\end{thebibliography}

\end{document}